\documentclass{elsart_e949}

\usepackage{epsfig}  
\usepackage{graphicx}
\usepackage{dcolumn}
\usepackage{bm}
\providecommand{\PGGUL}{\mbox {$8.3 \times 10^{-9}$\ }}      
\providecommand{\PGGULwo}{\mbox {$2.3 \times 10^{-8}$\ }}    
\providecommand{\PGGPSPAST}{\mbox {$5.0 \times 10^{-7}$\ }}  
\providecommand{\PGGPS}{\mbox {$6.0 \times 10^{-8}$\ }}      
\providecommand{\PGGPSIMPR}{\mbox {$8.3$\ }}                 
\providecommand{\PGLIM}{\mbox {$2.3 \times 10^{-9}$\ }}      
\providecommand{\KPITWO}{\mbox {$K_{\pi 2}$\ }}              
\providecommand{\NKCOUNT}{\mbox {$N_K=1.19 \times 10^{12}$\ }}  

\begin{document}

\begin{flushright}
\scriptsize
 {BNL/73917-2005-JA},
 {IHEP/2005-8},
 {KEK/2005-8},
 {TRIUMF/TRI-PP-05-07},
 {TUHEP/EX-05-001}
\normalsize
\end{flushright}

\begin{frontmatter}

\title{Search for the decay $K^+ \to \pi^+ \gamma \gamma$\\ in the $\pi^+$ momentum region $P>213$ MeV/$c$}

\collab{E949 Collaboration}

\author[ihep]{A.V.~Artamonov},
\author[unm]{B.~Bassalleck},
\author[bnl]{B.~Bhuyan\thanksref{bhu}},
\author[triumf]{E.W.~Blackmore},
\author[ubc]{D.A.~Bryman},
\author[triumf]{S.~Chen\thanksref{che}},
\author[bnl]{I-H.~Chiang},
\author[sbu]{I.-A.~Christidi},
\author[fnal]{P.S.~Cooper},
\author[bnl]{M.V.~Diwan},
\author[bnl]{J.S.~Frank},
\author[kyoto]{T.~Fujiwara},
\author[triumf]{J.~Hu},
\author[bnl]{D.E.~Jaffe},
\author[kek]{S.~Kabe},
\author[bnl]{S.H.~Kettell},
\author[inr]{M.M.~Khabibullin},
\author[inr]{A.N.~Khotjantsev},
\author[ualb]{P.~Kitching\thanksref{kit}},
\author[kek]{M.~Kobayashi},
\author[kek]{T.K.~Komatsubara},
\corauth[cor1]{{\em Email address:}}
\ead{takeshi.komatsubara@kek.jp (T.K.~Komatsubara).}
\author[triumf]{A.~Konaka},
\author[ihep]{A.P.~Kozhevnikov},
\author[inr]{Yu.G.~Kudenko},
\author[fnal]{A.~Kushnirenko\thanksref{kus}},
\author[ihep]{L.G.~Landsberg},
\author[unm]{B.~Lewis},
\author[bnl]{K.K.~Li},
\author[bnl]{L.S.~Littenberg},
\author[triumf]{J.A.~Macdonald\thanksref{mac}},
\author[triumf]{J.~Mildenberger},
\author[inr]{O.V.~Mineev},
\author[fukui]{M. Miyajima},
\author[kyoto]{K.~Mizouchi},
\author[ihep]{V.A.~Mukhin},
\author[rcnp]{N.~Muramatsu},
\author[rcnp]{T.~Nakano},
\author[lnso]{M.~Nomachi},
\author[kyoto]{T.~Nomura},
\author[triumf]{T.~Numao},
\author[ihep]{V.F.~Obraztsov},
\author[kek]{K.~Omata},
\author[ihep]{D.I.~Patalakha},
\author[ihep]{S.V.~Petrenko},
\author[triumf]{R.~Poutissou},
\author[fnal]{E.J.~Ramberg},
\author[bnl]{G.~Redlinger},
\author[kek]{T.~Sato},
\author[kek]{T.~Sekiguchi},
\author[nda]{T.~Shinkawa},
\author[bnl]{R.C.~Strand},
\author[kek]{S.~Sugimoto},
\author[fukui]{Y.~Tamagawa},
\author[fnal]{R.~Tschirhart},
\author[kek]{T.~Tsunemi\thanksref{tsu}},
\author[ihep]{D.V.~Vavilov},
\author[bnl]{B.~Viren},
\author[inr]{N.V.~Yershov},
\author[kek]{Y.~Yoshimura},
\author[kek]{T.~Yoshioka\thanksref{yos}}

\address[ihep]{Institute for High Energy Physics, Protvino, Moscow Region, 142 280, Russia}
\address[unm]{Department of Physics and Astronomy, University of New Mexico, Albuquerque, NM 87131, USA}
\address[bnl]{Brookhaven National Laboratory, Upton, NY 11973, USA}
\address[triumf]{TRIUMF, 4004 Wesbrook Mall, Vancouver, British Columbia, Canada V6T 2A3}
\address[ubc]{Department of Physics and Astronomy, University of British Columbia, Vancouver, British Columbia, Canada V6T 1Z1}
\address[sbu]{Department of Physics and Astronomy, Stony Brook University, Stony Brook, NY 11794, USA}
\address[fnal]{Fermi National Accelerator Laboratory, Batavia, IL 60510, USA}
\address[kyoto]{Department of Physics, Kyoto University, Sakyo-ku, Kyoto 606-8502, Japan}
\address[kek]{High Energy Accelerator Research Organization (KEK), Oho, Tsukuba, Ibaraki 305-0801, Japan}
\address[inr]{Institute for Nuclear Research RAS, 60 October Revolution Pr. 7a, 117312 Moscow, Russia}
\address[ualb]{Centre for Subatomic Research, University of Alberta, Edmonton, Canada T6G 2N5}
\address[fukui]{Department of Applied Physics, Fukui University, 3-9-1 Bunkyo, Fukui, Fukui 910-8507, Japan}
\address[rcnp]{Research Center for Nuclear Physics, Osaka University, 10-1 Mihogaoka, Ibaraki, Osaka 567-0047, Japan}
\address[lnso]{Laboratory of Nuclear Studies, Osaka University, 1-1 Machikaneyama, Toyonaka, Osaka 560-0043, Japan}
\address[nda]{Department of Applied Physics, National Defense Academy, Yokosuka, Kanagawa 239-8686, Japan}

\thanks[bhu]{Also at the Department of Physics and Astrophysics, University of Delhi, Delhi 110007, India. Present address: Department of Physics and Astronomy, University of Victoria, Victoria, British Columbia, Canada V8W 3P6.}
\thanks[che]{Present address: Department of Engineering Physics, Tsinghua University, Beijing 100084, P.R. China.}
\thanks[kit]{Present address: TRIUMF, Canada.}
\thanks[kus]{Present address: Institute for High Energy Physics, Protvino, Russia.}
\thanks[mac]{Deceased.}
\thanks[tsu]{Present address: Research Center for Nuclear Physics, Osaka University, Japan.}
\thanks[yos]{Present address: International Center for Elementary Particle Physics, University of Tokyo, Tokyo 113-0033, Japan.}

\begin{abstract}
 We have searched for the $K^+ \to \pi^+ \gamma \gamma$ decay
in the kinematic region with $\pi^+$ momentum close to the end point.
No events were observed, and the 90\% confidence-level upper limit on the 
partial branching ratio was obtained, 
${B}$($K^+ \to \pi^+ \gamma \gamma$, $P>213$ MeV/$c$) $<$ \PGGUL
under the assumption of chiral perturbation theory 
including next-to-leading order ``unitarity'' corrections.
The same data were used to determine an upper limit 
on the $K^+\to\pi^+\gamma$ branching ratio of \PGLIM 
at the 90\% confidence level.

\end{abstract}

\end{frontmatter}

\clearpage

We report the results of a search for the rare decay 
$K^+ \to \pi^+ \gamma \gamma$ 
in the $\pi^+$ momentum region $P>213$ MeV/$c$
from the E949 experiment~\cite{E949proposal} at
the Alternating Gradient Synchrotron (AGS) of 
Brookhaven National Laboratory.
The first observation of the decay 
in the $\pi^+$ momentum region 100--180 MeV/$c$
was reported~\cite{pggE787} by the E787 experiment at the AGS 
with a partial branching ratio of 
${B}$($K^+ \to \pi^+ \gamma \gamma$, 100 MeV/$c$ $<$ $P$ $<$180 MeV/$c$)
= $(6.0\ \pm\ 1.5(stat)\ \pm\ 0.7(syst))\times 10^{-7}$. 
In the region $P$ $>$215 MeV/$c$
no $K^+ \to \pi^+ \gamma \gamma$ decays were observed and,
assuming a pure phase-space kinematic distribution, 
a 90\% confidence-level (C.L.) upper limit of \PGGPSPAST  was set 
on the total branching ratio~\cite{pggE787}.
This established that the $K^+ \to \pi^+ \gamma \gamma$
background to the rare decay 
$K^+ \to \pi^+ \nu \bar\nu$~\cite{pnntheory}
was negligible.

\begin{figure}[h]
\begin{center}
\includegraphics*[width=0.70\linewidth]{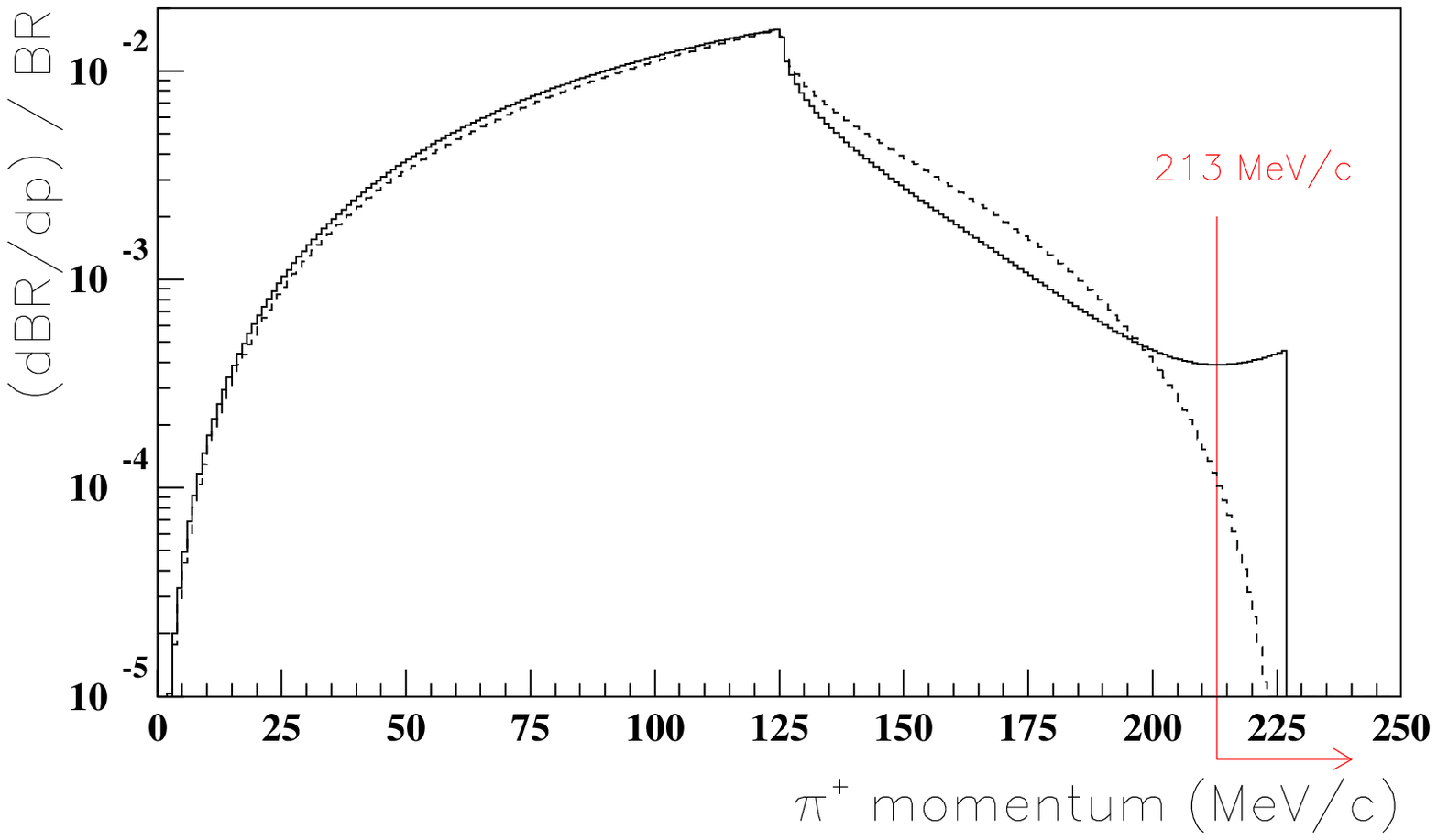}
\end{center}
\caption{Predictions for the $\pi^+$ momentum 
         for the ChPT parameter 
         {\it \^{c}} = $1.8$ including unitarity corrections
         (solid line) and 
         for {\it \^{c}} = $1.6$ without the corrections
         (dashed line) as described in the text.
         The ratios of the partial branching ratio
         in the kinematic region $P>213$ MeV/$c$, 
         indicated by the arrow, 
         to the total branching ratio are $5.77\times 10^{-3}$
         and $0.515\times 10^{-3}$,
         for {\it \^{c}} = $1.8$ and {\it \^{c}} = $1.6$ respectively.
} 
\label{fig:spectrum}
\end{figure}

In an effective-field approach to low-energy hadronic interactions
called chiral perturbation theory (ChPT)~\cite{ChPT}, 
there is no $O(p^2)$ (``tree-level'') contribution 
to $K^+ \to \pi^+ \gamma \gamma$ or its neutral counterpart 
$K^0_L \to \pi^0 \gamma \gamma$; 
the leading contributions start at $O(p^4)$~\cite{pggChPTo4}.
For $K^+ \to \pi^+ \gamma \gamma$, 
both the branching ratio and the $\pi^+$ spectrum shape
at $O(p^4)$
are sensitive to the undetermined coupling-constant {\it \^{c}}.
There is no complete calculation at the next-to-leading order,
$O(p^6)$. 
The dominant effects~\cite{pggChPTUC} are
one-loop ``unitarity'' corrections, 
deduced from an empirical fit 
to the decay amplitude of $K^+ \to \pi^+ \pi^+ \pi^-$
and containing the same constant {\it \^{c}},
and vector-meson exchange. 
In $K^+ \to \pi^+ \gamma \gamma$
vector-meson exchange
is expected to be negligible 
compared to unitarity corrections.
The corrections 
result in a slightly different prediction
        for the $\pi^+$ spectrum (Fig.~\ref{fig:spectrum}). 
Separate fits to the measured $\pi^+$ spectrum from \cite{pggE787}
with and without the unitarity corrections yielded
{\it \^{c}} = $1.8 \pm 0.6$ and {\it \^{c}} = $1.6 \pm 0.6$,
respectively, but slightly preferred their inclusion.
For $K^0_L \to \pi^0 \gamma \gamma$,  
the amplitude at $O(p^4)$ is determined
without any undetermined coupling-constant, 
but the measured branching ratio, 
$(1.41 \pm 0.12)\times 10^{-6}$~\cite{PDG2004},
is twice as large as predicted
at $O(p^4)$; 
the vector-meson contribution 
in the next-to-leading order calculation
(sometimes parametrized by an effective coupling constant $a_v$)
is considered to be important to this decay~\cite{pggChPTVMD}.

 One of the consequences of the unitarity corrections
to $K^+ \to \pi^+ \gamma \gamma$ 
is a non-zero amplitude in the kinematic region close to the end point 
of $P=$227 MeV/$c$,
 where the two-photon invariant mass $m_{\gamma\gamma}=$0 MeV/$c^2$, 
as shown in Fig.~\ref{fig:spectrum}.
The partial branching ratio 
${B}$($K^+ \to \pi^+ \gamma \gamma$, $P>213$ MeV/$c$),
corresponding to $m_{\gamma\gamma}<$108 MeV/$c^2$, 
is predicted to be 
$6.10 ^{+0.16}_{-0.12} \times 10^{-9}$
         for {\it \^{c}} = $1.8 \pm 0.6$ including unitarity corrections
and
$0.49 ^{+0.23}_{-0.18} \times 10^{-9}$
         for {\it \^{c}} = $1.6 \pm 0.6$ without the corrections.
Observation of the decay at a partial branching ratio larger than predicted
would indicate the contribution of vector-meson exchange
or other new dynamics to $K^+ \to \pi^+ \gamma \gamma$.
The kinematic region close to the end point in $K^0_L \to \pi^0 \gamma \gamma$
is known to be crucial to understand the CP-conserving component
to the $K^0_L \to \pi^0 e^+ e^-$ decay
through the 
$K^0_L \to \pi^0 \gamma \gamma \to \pi^0 e^+ e^-$ amplitude, but 
experimental results on $a_v$ 
in $K^0_L \to \pi^0 \gamma \gamma$~\cite{pggKTeV,pggNA48} 
are inconsistent
(see Ref.~\cite{pggGV,pggpi0ee} for theoretical discussions).

\begin{figure}[h]
\begin{center}
\includegraphics*[width=.45\linewidth]{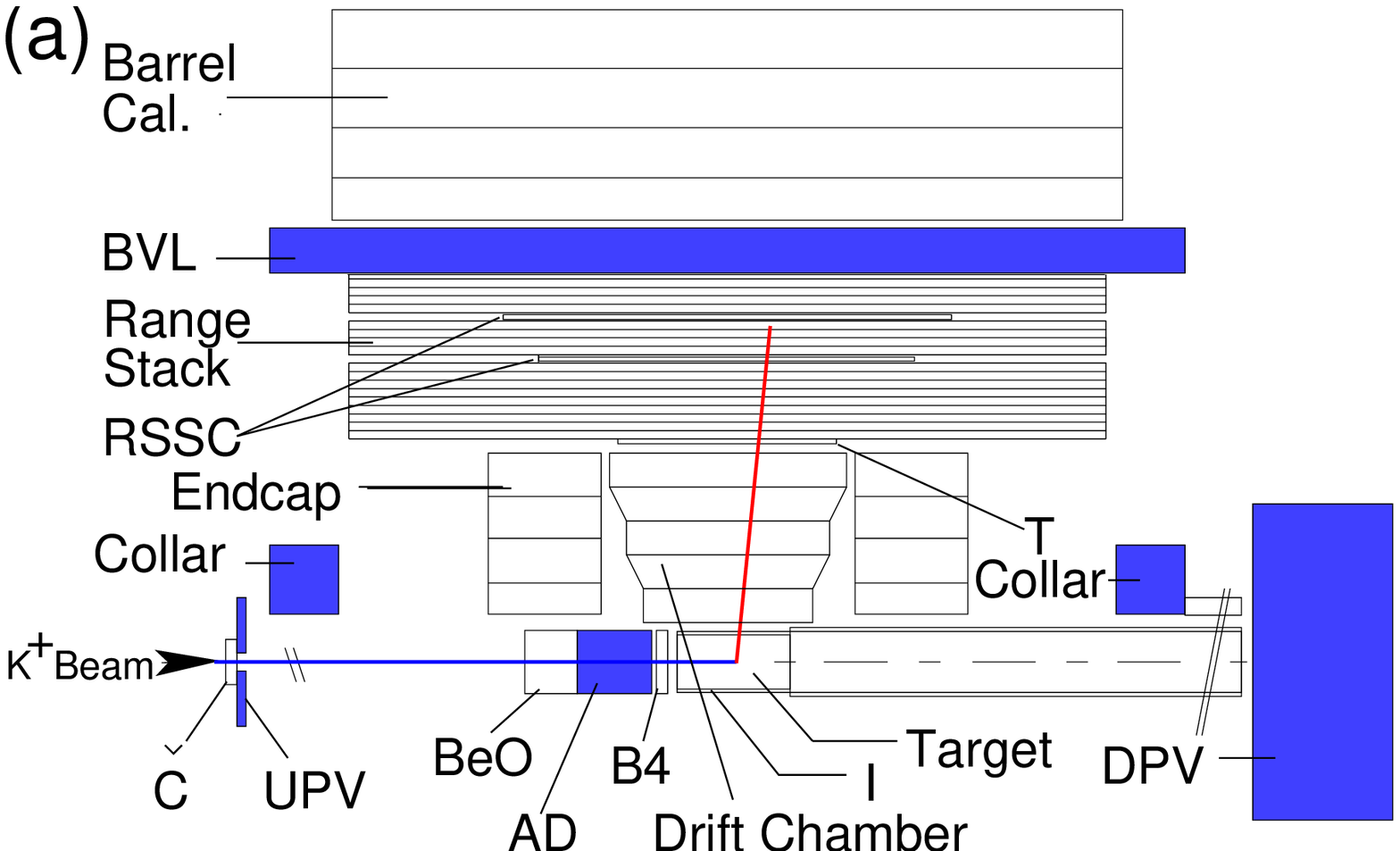}
\includegraphics*[width=.45\linewidth]{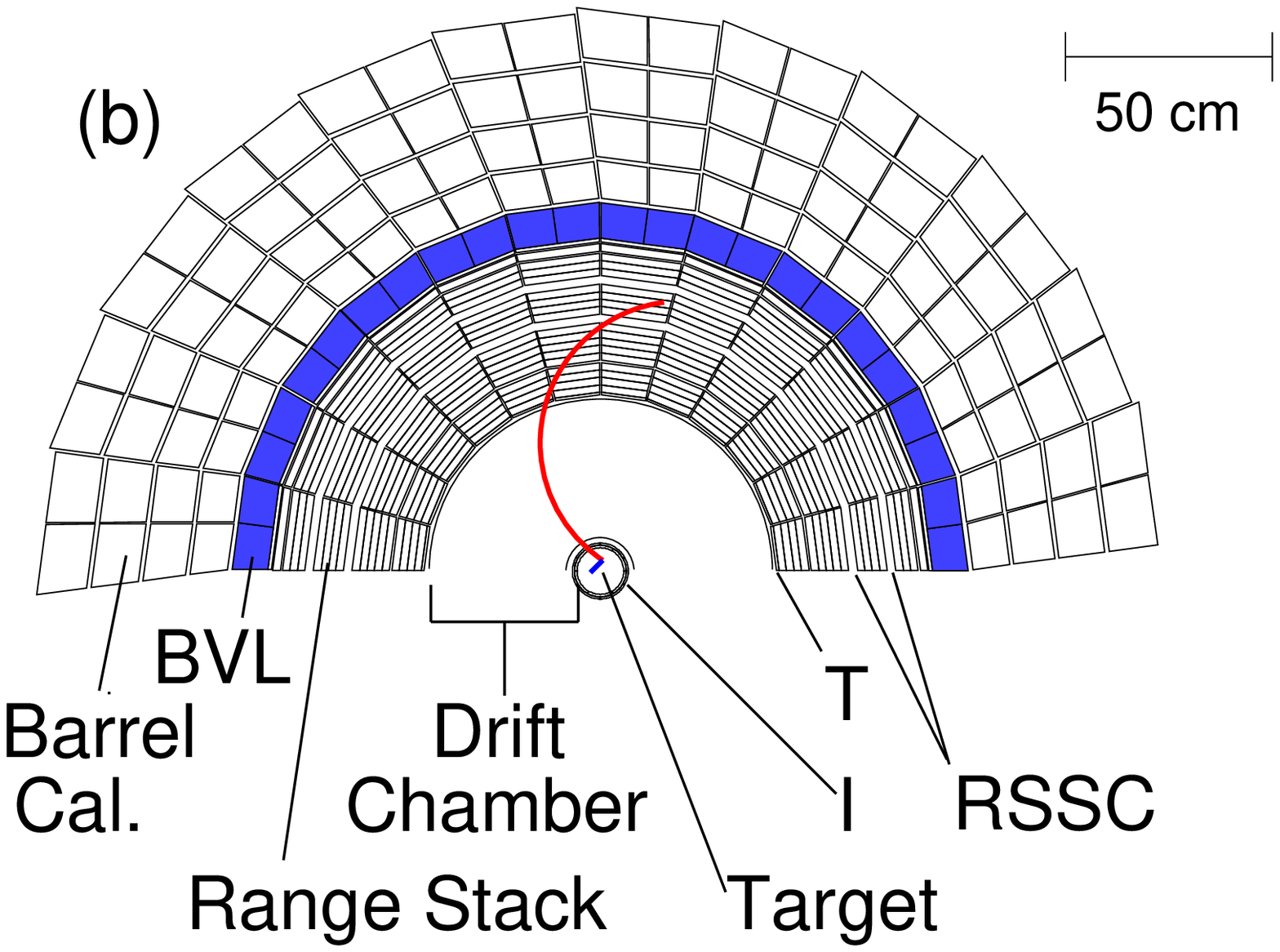}
\end{center}
\caption{Schematic side (a) and end (b) views of the upper half
of the E949 detector. 
\v{C}: \v{C}erenkov counter;
B4: energy-loss counters; 
I and T: inner and outer trigger scintillation counters; 
RSSC: RS straw-tube tracking chambers. 
New or upgraded subsystems for E949 
 (shaded) included  the
barrel veto liner (BVL), collar,
upstream photon-veto (UPV), 
active degrader (AD), and downstream photon-veto (DPV).}
\label{fig:detector}
\end{figure}

 E949 was designed primarily to measure the decay 
 $K^+ \to \pi^+ \nu \bar\nu$~\cite{pnnE949}.
 The AGS delivered kaons of 710 MeV/$c$ to the experiment
 at a rate of $12.8 \times 10^{6}$ per 2.2-s spill. 
 Kaons, detected and identified by \v{C}erenkov, tracking, and energy-loss counters,
 were slowed by BeO and active degraders, and 
 came to rest and decayed
 in a scintillating-fiber target.
  The exposure of kaons in each spill was recorded by a scaler 
  which counted the number of kaons entering the target during the time 
  the data acquisition system was ready to accept data.
 Fig.~\ref{fig:detector} shows a diagram of the apparatus.  
 Measurements of charged
 decay products were made using the target, a central drift chamber,
 and a cylindrical range stack (RS) composed of 18 layers of plastic
 scintillator with two embedded layers of tracking chambers. 
 The RS counters in the first layer (T-counters in Fig.~\ref{fig:detector})
 were 0.635-cm thick and 52-cm long; 
 the subsequent RS counters were 1.905-cm cm thick and 182-cm long.   
 The pion from the $K^+ \to \pi^+ \gamma \gamma$ decay 
 was identified by observation of the 
 $\pi^+ \!  \rightarrow \! \mu^+ \!  \rightarrow \!  e^+$ 
 decay sequence in the RS 
 using 500-MHz waveform digitizers
 based on flash analogue-to-digital converters~\cite{detTD}. 
 Trigger counters surrounding the target 
 (I-counters in Fig.~\ref{fig:detector} which were 0.64-cm thick)  
 and the T-counters surrounding the drift chamber
 defined the fiducial region. 
 Counters (0.95-cm thick) surrounding the RS
 helped to suppressed the muons from $K^+ \to \mu^+ \nu$ and
 $K^+ \to \mu^+ \nu \gamma$ decays
 by identifying long-range muons 
 that had completely traversed the RS.
 A hermetic calorimeter system surrounded the central region.
 The photons from $K^+ \to \pi^+ \gamma \gamma$
 were detected in a lead/scintillator sandwich barrel detector (BV)
 surrounding the RS, while
 two endcap calorimeters 
 and other subsystems (collar, ``UPV'', ``AD'', and ``DPV'' in Fig.~\ref{fig:detector})  
 were used for detecting extra particles. 
 A  solenoid   surrounding the BV provided
 a 1 T  magnetic field along the beam line.

The AGS proton-beam intensity, and the E949 beam line
and apparatus were improved
over those used in E787~\cite{E787det}
for the $K^+ \to \pi^+ \gamma \gamma$ study, 
which was performed in 1991\footnote{
  Many of the improvements were made in the E787 apparatus after 1991.}.
The new beam line~\cite{LESB3} incorporated two stages of electrostatic 
particle separation which improved the acceptance for kaons 
as well as the $K^+$ to $\pi^+$ ratio.
The target, central drift chamber, and RS tracking chambers were
replaced by a new target consisting of 0.5-cm square fibers,
a new low-mass drift chamber~\cite{detUTC}, and straw-tube chambers, 
respectively. 
One third of the RS scintillation counters were
replaced to  increase the  light output.
A new photon detector, the barrel veto liner (BVL), was installed 
to add 2.3 radiation lengths of lead/scintillator
  sandwich material to the BV.
The endcaps were replaced by new fully active detectors
consisting of undoped-CsI crystals~\cite{detCsI}
with significantly increased light output,
and both the target and endcaps were read out 
using 500-MHz waveform digitizers
based on charged coupled devices~\cite{detCCD}
to improve timing and double-pulse resolution.
Additional ancillary photon-veto systems~\cite{detCollar} and a flasher system of Light Emitting Diode
to aid in the RS energy calibration were also introduced.

The 2002 data set used for this analysis derived from a
total exposure of kaons entering the target \NKCOUNT .
The trigger required a kaon at rest in the target to
decay at least 1.5 ns later into a positively charged
particle which subsequently came to rest in the RS, 
accompanied by
coincident electromagnetic showers in both the BVL and BV, 
and no extra energy in the endcap or RS counters.
The decay particle was required to penetrate the RS to at
least the 16th layer to suppress backgrounds from the
monochromatic $K^+ \to \pi^+ \pi^0$ decay (\KPITWO) with P=205~MeV/c, 
and no further than the 17th layer to suppress decays into muons.
In the RS counter where the particle came to rest,
called the ``stopping counter'',
a $\pi^+ \!  \rightarrow \! \mu^+$ decay was identified
online based on the pulse shape information 
from the waveform digitizers on the RS.
An improved trigger system~\cite{L0board}
including a programmable trigger board
reduced the online dead time.
A total of $1.1 \times 10^7$ events met the trigger requirements. 

The momentum, equivalent range in plastic scintillator ($R$),
  and kinetic energy ($E$) of the charged track were 
  reconstructed with information from the target, drift chamber and RS.
Tracks were accepted for the region defined by 
  213 MeV/$c$ $< P <$ 234 MeV/$c$, 
  33.5 cm $< R <$ 41.3 cm, 
  and 116 MeV $< E <$ 135 MeV,
 where the lower limits corresponded to 3.3, 2.3, and 2.6 standard deviations,
 respectively, above the \KPITWO\ peak ($P=205$ MeV/$c$, $R=30.4$ cm, and $E=108$ MeV).
The larger search region ($P>$ 213 MeV/$c$)
compared with E787 ($P>$ 215 MeV/$c$)
resulted from improved kinematic reconstruction, which also
removed the requirement of 
a constrained fit for consistency with $K^+ \to \pi^+ \gamma \gamma$
kinematics that was used in E787~\cite{pggE787}.

  The timing and energy ($E_{\gamma}$) of the photons 
  were determined by grouping  adjacent hit modules in the BVL and BV 
 to classify isolated photon showers (``clusters'').
  The hit position in each module along the beam axis ($z$) was calculated
 from the end-to-end time and energy differences; 
  the azimuthal angle ($\phi$) of the hit position 
 in the end view of the detector was determined
   by the segmentation of the modules.
  The location of the photon shower in $z$ and $\phi$ was obtained 
  by an energy-weighted average of the hit positions and was used, 
  in conjunction with the kaon-decay vertex position in the target, 
  to determine the polar and azimuthal angles of the photon
  as well as
  the opening angles between the photon and the charged track 
  in the side view ($\theta_{\pi^+\gamma}$)
  and in the end view ($\phi_{\pi^+\gamma}$).
  In approximately half of the 
  $K^+ \to \pi^+ \gamma \gamma$ decays
  the two photons were unresolved and appear as a single cluster 
  in the BVL and BV, 
  since the opening angle between two photons from 
  $K^+ \to \pi^+ \gamma \gamma$ gets smaller for the events 
  with $\pi^+$ momentum close to the kinematic end point.
  The events with either one or two clusters were accepted; 
  at least one cluster with 50 MeV $< E_{\gamma}<$ 320 MeV, 
  $\theta_{\pi^+\gamma} >$ 155$^{\circ}$, and $\phi_{\pi^+\gamma} >$ 155$^{\circ}$
  was required.
  For events with two clusters,  
  the lower-energy cluster had to have $E_{\gamma}>$ 10 MeV. 

  Background sources from kaon decays at rest were classified 
  into three types:
  \begin{description} 
    \item[``$mismeasured$'']:
        \KPITWO decays with mismeasurements of the $\pi^+$
         and the two photons,  
    \item[``$overlap$'']:
        \KPITWO decays with the
        lower-energy photon overlapping the $\pi^+$ track in the RS,
        causing the kinetic energy of reconstructed $\pi^+$ track to be
        incorrectly measured, and
    \item[``$muon$'']: 
         $K^+$ decays with a $\mu^+$ misidentified as a $\pi^+$ and
         with photons in the final state 
         (e.g. $K^+ \to \mu^+ \nu \gamma$, $K^+ \to \pi^0 \mu^+ \nu$ and 
          \KPITWO with  $\pi^+$ decay in flight). 
  \end{description}
  Another background source was due to kaon decay in flight:
  \begin{description}
    \item[``$DIF$'']:
           \KPITWO decay-in-flight before the kaon came to rest in the target.
  \end{description}
  Beam-related backgrounds (e.g. multiple beam particles scattering into the
  detector) were found to negligible.

 These backgrounds were studied from the data 
 by imposing offline selection criteria (``cuts'').
 The requirements on the $\pi^+$ momentum, range and
kinetic energy provided large suppression of all of the backgrounds from
kaon decays at rest.
Several cuts, referred to as ``$\gamma$ selection cuts'', 
were imposed to suppress the $mismeasured$ background.
These included cuts on the following quantities: 
 \begin{itemize}
   \item the cut on the invariant mass of two photons
         to reject events with two clusters 
         and $m_{\gamma\gamma}>$100 MeV/$c^2$,
   \item the cut on additional coincident energy  
         to reject events with 
           activity not associated with the $\pi^+$ and 
           the candidate signal photons\footnote{
              The extra activity was identified in the various subsystems, 
              including the RS, 
              as hits in the counters in coincidence with the $\pi^+$ track
              within a few ns and with energy above a threshold 
              of typically 1 MeV.}, and
    \item the cut on the photon clusters
          to reject events with 
          two photons from a $\pi^0$ which hit the same or adjacent  
          modules in BVL and BV and form a single cluster\footnote{
                  Due to the kinematics of \KPITWO and 
                  subsequent $\pi^0\to\gamma\gamma$ decays,
                  the two photons must hit the modules at different $z$ positions 
                  along the beam axis.
                  An event was rejected 
                  if the maximum discrepancy among the $z$-position measurements in the 
                   modules of the cluster was larger than 113 cm.}~\cite{E787pig}.
 \end{itemize}
 In addition, only those events whose total photon-shower energy was deposited 
 in the BVL and BV calorimeters were accepted.
 Cuts on $dE/dx$ in the RS
 to reject events with a RS counter
    in which the measured energy was larger than expected 
     from the reconstructed range in that counter
 were imposed to suppress the $overlap$ background~\cite{E787pig}. 
 The cuts on the relation between the range measured in the RS 
 and the momentum  measured in the drift chamber 
 as well as  
 the cuts on the  $\pi^+ \!  \rightarrow \! \mu^+ \!  \rightarrow \!  e^+$ 
 decay sequence, recorded in the RS stopping counter,
 were also imposed to suppress the $muon$ background.
 The $DIF$ background was suppressed
  by requiring a delay of at least 2 ns between
  the time of the kaon coming to rest in the target 
   and its subsequent decay,  
  and by imposing the cuts on the timing 
  between the $\pi^+$ in the RS and 
  the $K^+$ in the \v{C}erenkov counter.

 To study and measure these backgrounds,
 two independent sets of cuts were established for each background source.
 At least one of these cuts was inverted 
 to enhance each background 
 collected by the $K^+ \to \pi^+ \gamma \gamma$ trigger 
 as well as to prevent candidate events from being examined 
 before the background studies were completed.
 To avoid contamination from other background sources, 
 all the cuts except for those being studied were 
 imposed on the data.
  Possibilities of a correlation between the two sets of cuts or of a
  biased estimate of the effectiveness of the cuts were studied,
  and were found to be negligible. 
  The signal acceptance and background levels were studied
  as a function of cut severity.
  A comparison of the
  observed background levels near but outside the signal region was made 
  to the predicted background in these regions.
  All of these studies~\cite{ThYoshioka} were performed 
  in the same manner as the $K^+ \to \pi^+ \nu \bar\nu$ analyses of 
  E787~\cite{pnnE787} and E949~\cite{pnnE949}.
 Table~\ref{tab:background} summarizes 
 the background levels measured with the final analysis cuts
 and the two sets of cuts for studying each background source. 
 In total, $0.197\pm 0.070$ background events were expected in the signal region.

 The acceptance ($A$) and the single event sensitivity ($SES$)
 for  $K^+ \to  \pi^+ \gamma\gamma$
 in the kinematic region $P>213$ MeV/$c$
 were derived from the acceptance factors in Table~\ref{tab:acceptance}
 and the total kaon exposure \NKCOUNT 
 times the $K^+$-stopping efficiency,
 which is the fraction of kaons entering the target
 that came to rest before decaying.
 The efficiency was measured  to be $0.754\pm 0.024$
 with the \KPITWO events  collected 
 with the $K^+ \to \pi^+ \gamma \gamma$ trigger
 and selected using all the analysis cuts
 except those designed to remove \KPITWO events.
 Thus the sensitivity
 for  $K^+ \to  \pi^+ \gamma\gamma$ was normalized to \KPITWO and 
 many systematic uncertainties in the measurement of 
 the acceptance factors, in particular in the trigger acceptance, 
 canceled.
 We obtained  
 $A=( 2.99\pm 0.07 )\times 10^{-4}$ 
 and $SES=( 3.72\pm 0.14) \times 10^{-9}$ 
           for {\it \^{c}} = 1.8 including unitarity corrections
 and 
 $A=( 1.10\pm 0.04 )\times 10^{-4}$
 and $SES=( 1.01\pm 0.05) \times 10^{-8}$
          for {\it \^{c}} = 1.6 without the corrections.
 The former sensitivity was
 below the predicted branching ratio of $6.10 \times 10^{-9}$,
 giving an expectation of 1.6 events.
 In order to verify that the sensitivity estimations were correct,
 a sample of $K^+ \to \mu^+ \nu$ decays
 accumulated with a  calibration trigger
 was analyzed. 
 The branching ratio of
 $0.628\pm 0.020$
 measured with the same $K^+$-stopping efficiency
 was  consistent with the Particle Data Group value~\cite{PDG2004}.
 The systematic uncertainty in the sensitivity was estimated to be
 less than 10\%.

\begin{figure}[h]
\begin{center}
\includegraphics*[width=.70\linewidth]{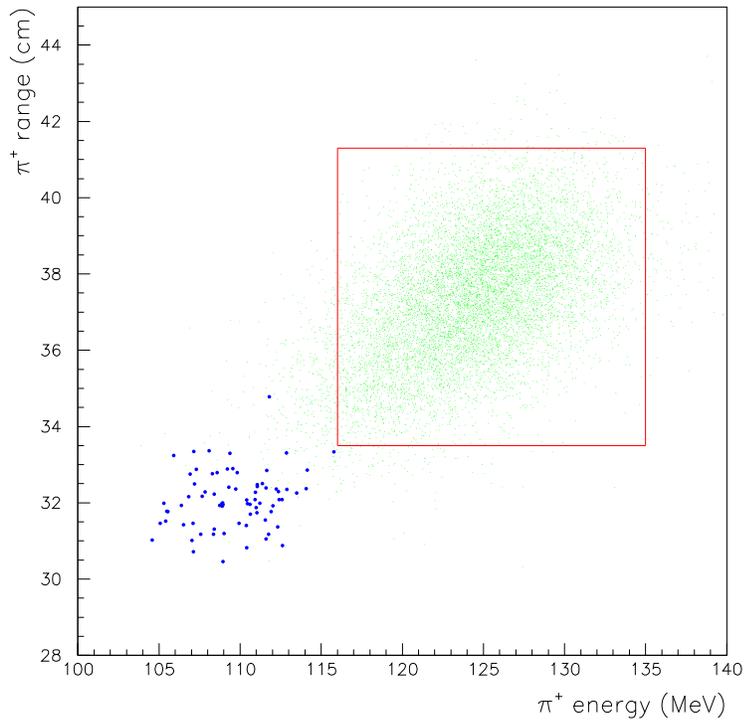}
\end{center}
\caption{Range (equivalent cm of plastic scintillator) vs. kinetic energy (MeV) plot 
 of the events with all analysis cuts imposed. 
 The box indicates the signal region for 
 the $K^+ \to \pi^+ \gamma \gamma$ decay. 
 The dark points represent the data. 
 The simulated distribution of
 expected events from $K^+ \to \pi^+ \gamma\gamma$
  for {\it \^{c}} = $1.8$ including unitarity corrections
 is indicated by the light dots.}
\label{fig:boxpigg}
\end{figure}

 After imposing all analysis cuts, 
 no events were observed in the signal region (Fig.~\ref{fig:boxpigg}).
 The group of 74 events around 
 $R=32$ cm and $E=110$ MeV
 are due to the \KPITWO background.
 Taking 2.24 events instead of zero 
 according to the unified approach~\cite{F-C}
 with the background contribution of 0.197 events, 
 we set a 90\% C.L. upper limit 
 on the partial branching ratio
 ${B}$($K^+ \to \pi^+ \gamma \gamma$, $P>213$ MeV/$c$)
 as
 \PGGUL
            for {\it \^{c}} = 1.8 including unitarity corrections 
 and
 \PGGULwo
           for {\it \^{c}} = 1.6 without the corrections.
 The systematic uncertainty was not taken into consideration 
 in deriving the limits.
 For the purpose of comparison with the previous E787 results,
 a 90\% C.L. upper limit for the total $K^+ \to  \pi^+ \gamma\gamma$
 branching ratio assuming the phase-space distribution 
 was calculated; the present limit \PGGPS
 is \PGGPSIMPR times lower than 
 the same limit in E787 (\PGGPSPAST).

 The data described above were also used to set an upper limit
 on the branching ratio for $K^+\to\pi^+\gamma$ decay, 
 which is forbidden by angular-momentum 
 conservation and gauge invariance, but is allowed 
 in noncommutative theories~\cite{NCSM}.
 The signature of $K^+\to\pi^+\gamma$
 was a two-body decay of a kaon at rest with a 227-MeV/$c$ $\pi^+$ track
 in the RS and a 227-MeV photon emitted directly opposite to it
 and observed in the BVL and BV calorimeters.
  The trigger, event reconstruction, and offline selection criteria
  in the study of $K^+ \to  \pi^+ \gamma\gamma$ had been designed 
  so that the same data were available to the search for 
  $K^+ \to  \pi^+ \gamma$.
  Since the background levels were already small,
  the $\pi^+$ accepted region was not reduced 
  for the $K^+ \to  \pi^+ \gamma$ analysis.
 The previous limit from the E787 study was 
 $3.6\times 10^{-7}$~\cite{E787pig} (90\% C.L.)
 from data collected in 1996--1997 with 
 a highly prescaled trigger with relaxed trigger-conditions
 resulting in the total kaon exposure of $6.7\times 10^{8}$.
 The new 90\% C.L. upper limit from E949, 
 using the acceptance for $K^+\to\pi^+\gamma$ of
  $( 1.08\pm 0.02 )\times 10^{-3}$,  is 
\PGLIM .

 The results from this study cannot confirm nor rule out  
 the unitarity corrections of ChPT, but
 the upper limits obtained are the the most restrictive yet achieved on
  $K^+ \to \pi^+ \gamma\gamma$ 
  and as well as on $K^+\to\pi^+\gamma$.
The E949 experiment has been shown to be suitable for the
study of $K^+ \to \pi^+ \gamma\gamma$ in the $\pi^+$ momentum region
close to the end point. The experimental uncertainty is
limited by statistics; additional data is required to more
stringently test the predictions of ChPT for this decay.
 The possibility to observe the $K^+ \to \pi^+ \gamma\gamma$ decay
 in the kinematic region, 
 if the ChPT including unitarity corrections is correct, 
 gives further impetus for additional data collection.

\begin{ack}

We gratefully
acknowledge the dedicated effort of the technical staff supporting
E949 and of the Brookhaven C-A Department.
We are also grateful to 
 G.~D'Ambrosio, 
 F.~Gabbiani,
 G.~Isidori, 
 and G.~Valencia
for useful discussions. 
This research
was supported in part by the U.S. Department of Energy, 
the Ministry of Education, Culture, Sports, Science and Technology of
Japan through the Japan-U.S. Cooperative Research Program in High
Energy Physics and under Grant-in-Aids for Scientific Research, the
Natural Sciences and Engineering Research Council and the National
Research Council of Canada, the Russian Federation 
State Scientific Center Institute for High
Energy Physics, and the Ministry of Science and Education 
of the Russian Federation.

\end{ack}

\clearpage


\clearpage

\clearpage

\begin{table}[h]
\begin{center}
\begin{tabular}{|ll||l|l|} 
\hline
\multicolumn{1}{|l}{~Source}       &  background level\ \  & \multicolumn{2}{c|}{two sets of cuts} \\
\hline
\hline
~$mismeasured$\ \ \ & $0.017\pm 0.006$      &\ \ $\pi^+$ accepted region ($P$,$R$,$E$)\ \  &\ \ $\gamma$ selection\ \  \\
\hline
~$overlap$          & $0.065\pm 0.065$      &\ \ $\pi^+$ accepted region ($P$,$R$)\ \      &\ \ $\pi^+$ $dE/dx$\ \ \\
\hline
~$muon$             & $0.090\pm 0.020$      &\ \ $\pi^+$ accepted region ($P$,$R$,$E$)\ \  &  
                                           \ \ $\pi^+ \! \rightarrow \! \mu^+ \! \rightarrow \! e^+$\ \ \\
                  &                       &\ \ $\pi^+$ range-momentum relation\ \               &    \\
\hline
~$DIF$              & $0.025\pm 0.014$      &\ \ delay in the target\ \  &
                                                        \ \ RS - \v{C} timing\ \ \\
\hline
\end{tabular}
\end{center}
\vspace*{0.5cm}
\caption{Expected background levels in the signal region
         and the two sets of cuts for studying each background source.
         The total background was $0.197\pm 0.070$ events.}
\label{tab:background}
\end{table}

\clearpage

\begin{table}[h]
\begin{center}
\begin{tabular}{llll} 
  Acceptance factors             		& UC            & w/o UC
	& samples\\
\hline
 Trigger
                                                &  $0.0623$     & $0.0407$
	&MC, $K_{B}$\\ 
 $\pi^+$ reconstruction and fiducial cuts				
						&  $0.980$	& $0.935$
	&MC, $K_{\mu 2}$\\ 
 $\pi^+$ accepted region ($P$,$R$,$E$)
						&  $0.912$	& $0.667$
	&MC\\
 $\pi^+$ stop without nuclear interaction\ \ \ \                      
						&  $0.492$	& $0.524$
	&MC\\
 \ \ or decay-in-flight
						&   		&
	&\\
 $dE/dx$ and kinematic cuts
						&  $0.537$	& $0.537$
	&$K_{\pi 2}$, $\pi_{scat}$\\
 $\pi^+ \! \rightarrow \! \mu^+ \! \rightarrow \! e^+$ cuts 
						&  $0.349$	& $0.349$
	&$\pi_{scat}$\\
 $\gamma$ reconstruction and fiducial cuts
						&  $0.530$	& $0.492$
	&MC, $K_{\pi 2}$\\
 $\gamma$ selection cuts
						&  $0.216$	& $0.177$
	&MC, $K_{\pi 2}$\\
 Other cuts on beam and target 		
						&  $0.507$	& $0.507$
	&$K_{\mu 2}$\\

\hline
Total acceptance
                                                &  $2.99\times 10^{-4}$\ \ \ \ 	& $1.10\times 10^{-4}$\ \ \ \ 
 	&\\
\end{tabular}
\end{center}
\vspace*{0.5cm}
\caption{Acceptance factors
         for the $K^+ \to  \pi^+ \gamma\gamma$ decay
         in the kinematic region $P>213$ MeV/$c$, 
         for {\it \^{c}} = 1.8 including unitarity corrections
         (``UC'')
         and 
         for {\it \^{c}} = 1.6 without the corrections (``w/o UC''), 
         and the samples used to determine them. 
         ``MC'' in the rightmost column means the sample generated by Monte Carlo simulation.
         ``$K_{B}$'', ``$K_{\mu 2}$'', ``$K_{\pi 2}$'', and ``$\pi_{scat}$'' mean
         the data samples of kaons entering the target, 
         $K^+ \to \mu^+ \nu$ decays, \KPITWO  decays, 
         and scattered beam pions,
         respectively; 
         these samples were accumulated by calibration triggers simultaneous to the collection
         of signal candidates.}
\label{tab:acceptance}
\end{table}


\begin{thebibliography}{ (a) }


\bibitem{E949proposal}
 \mbox{B. ~Bassalleck {\it et al.}}, E949 Proposal, BNL-67247, TRI-PP-00-06 (1999),
 [http://www.phy.bnl.gov/e949/] .

\bibitem{pggE787}
 P.~Kitching {\it et al.}, 
  Phys. Rev. Lett. {79} (1997) 4079.

\bibitem{pnntheory}
 A.J.~Buras, F.~Schwab, and  S.~Uhlig, 
  hep-ph/0405132,
  and references therein.

\bibitem{ChPT} 
 J.F.~Donoghue, E.~Golowich, and B.R.~Holstein, 
 {\it Dynamics of the Standard Model}
 (Cambridge University Press, Cambridge, 1992),  
 and references therein.

\bibitem{pggChPTo4}
 G.~Ecker, A.~Pich, and E.~de Rafael, 
   Phys. Lett. {B 189} (1987) 363; 
   Nucl. Phys. {B 303} (1988) 665;
 L.~Cappiello and G.~D'Ambrosio, 
   Nuovo Cimento {A 99} (1988) 155. 

\bibitem{pggChPTUC}
 G.~D'Ambrosio and J.~Portol\'{e}s, 
   Phys. Lett. {B 389} (1996) 770;
   Nucl. Phys. {B 492} (1997) 417;
   Nucl. Phys. {B 533} (1998) 494.
     

\bibitem{PDG2004}
 Particle Data Group, S.~Eidelman {\it et al.}, 
   Phys. Lett. {B 592} (2004) 1.

\bibitem{pggChPTVMD}
 G.~Ecker, A.~Pich, and E.~de Rafael, 
   Phys. Lett. {B 237} (1990) 481;
 L.~Cappiello, G.~D'Ambrosio, and M.~Miragliuolo, 
   Phys. Lett. {B 298} (1993) 423;
 A.G.~Cohen, G.~Ecker, and A.~Pich,  
   Phys. Lett. {B 304} (1993) 347.


\bibitem{pggKTeV}
 A.~Alavi-Harati {\it et al.},
  Phys. Rev. Lett. {83} (1999) 917.

\bibitem{pggNA48}
 A.~Lai {\it et al.},
   Phys. Lett. {B 536} (2002) 229.

\bibitem{pggGV}
 F.~Gabbiani and G.~Valencia, 
  Phys. Rev. {D 64} (2001) 094008;
  Phys. Rev. {D 66} (2002) 074006.

\bibitem{pggpi0ee}
 G.~Buchalla, G.~D'Ambrosio, and G.~Isidori, 
   Nucl. Phys. {B 672} (2003) 387.

\bibitem{pnnE949}
 V.V.~Anisimovsky {\it et al.}, 
  Phys. Rev. Lett. {93} (2004) 031801.

\bibitem{detTD}
M.S.~Atiya {\it et al.}, Nucl. Instrum. Methods Phys. Res., Sect. A {279} (1989) 180.

\bibitem{E787det}
 M.S.~Atiya {\it et al.}, 
  Nucl. Instrum. Methods Phys. Res., Sect. A {321} (1992) 129.

\bibitem{LESB3}
 J.~Doornbos {\it et al.}, 
  Nucl. Instrum. Methods Phys. Res., Sect. A {444} (2000) 546.

\bibitem{detUTC}
 E.W.~Blackmore {\it et al.}, 
  Nucl. Instrum. Methods Phys. Res., Sect. A {404} (1998) 295.

\bibitem{detCsI}
 I-H.~Chiang {\it et al.}, 
  IEEE Trans. Nucl. Sci. {42} (1995) 394;
 T.K.~Komatsubara {\it et al.}, 
  Nucl. Instrum. Methods Phys. Res., Sect. A {404} (1998) 315.

\bibitem{detCCD}
 D.A.~Bryman {\it et al.},  
  Nucl. Instrum. Methods Phys. Res., Sect A  {396} (1997) 394.


\bibitem{detCollar}
 O.~Mineev {\it et al.},
  Nucl. Instrum. methods Phys. Res., Sect. A {494} (2002) 362.

\bibitem{L0board}
 T.~Yoshioka {\it et al.}, 
  IEEE Trans. Nucl. Sci. {51} (2004) 334.


\bibitem{E787pig}
 S.~Adler {\it et al.}, 
  Phys. Rev. {D 65} (2002) 052009.

\bibitem{ThYoshioka}
  T.~Yoshioka,
  Ph.D. thesis, University of Tokyo, 
  2005.



\bibitem{pnnE787}
 S.~Adler {\it et al.}, 
   Phys. Rev. Lett. {88} (2002) 041803;
 S.~Adler {\it et al.}, 
   Phys. Rev. Lett. {84} (2000) 3768;
 S.~Adler {\it et al.}, 
    Phys. Rev. Lett. {79} (1997) 2204.



\bibitem{F-C}
 G.J.~Feldman and R.D.~Cousins, 
  Phys. Rev. {D 57} (1998) 3873.

\bibitem{NCSM}
  J.~Trampeti\'{c}, hep-ph/0212309; 
  B.~Meli\'{c}, K.~Passek-Kumeri\v{c}ki, and J.~Trampeti\'{c},
        hep-ph/0507231.


\end{thebibliography}
\end{document}